\documentclass[onecolumn,aps,preprint,floatfix]{revtex4}
\usepackage{graphicx}
\usepackage{amsmath,amssymb,amsfonts,mathrsfs}
\begin{document}
\author{M.~Sega}
\author{E.~Autieri}
\author{F.~Pederiva}
\affiliation{Dipartimento di Fisica and I.N.F.N., Universit\`a degli Studi di Trento, via Sommarive 14, 38100 Trento, Italy}
\begin{abstract}
Cremer--Pople puckering coordinates appear to be the natural candidate variables
to  explore the conformational space of cyclic compounds, and in literature different
parametrizations have been used to this end. However, while every parametrization
is equivalent in identifying conformations, it is not obvious that they can also
act as proper collective variables for the exploration of the puckered conformations 
free energy surface. It is shown that only the polar parametrization
is fit to produce an unbiased estimate of the free energy landscape. As
an example, the case of a six-membered ring, glucuronic acid, is presented,
showing the artefacts that are generated when a wrong parametrization is used.

\end{abstract}
\title{On the Calculation of Puckering Free Energy Surfaces}
\maketitle
\section{Introduction}
Cyclic and heterocyclic compounds plays a particularly relevant role in many
chemical and biological processes. Carbohydrates, for example, are one of the
fundamental building blocks of the biochemical structure and activity of the cell,
including energy transport, cell recognition and signaling. Knowing the structural
and conformational properties of cyclic compounds is therefore a task of primary
interest. The quantitative description of puckered conformation of $N$-membered
rings is of fundamental importance for the physics and chemistry of cyclic compounds.

After early attempts to rigorously describe puckered forms, starting from the
work by Kilpatrik\cite{Kilpatrick47}, the problem of a general definition of
puckering coordinates was eventually settled by Cremer and Pople\cite{Cremer75}.
A ring conformation is uniquely identified by just $N-3$ parameters, and is actually
representative of an infinite set of points in the $(3N-6)$-dimensional space of
configurations. Although puckering coordinates apply to rings of an arbitrary
number of members, their interpretation is already not straightforward for
$N=7$~\cite{Boessenkool80}, and soon becomes a daunting task for larger
$N$~\cite{Evans88,Evans90} (choices other than the Cremer--Pople one are
possible\cite{Strauss71}, although completely equivalent\cite{Boeyens89}).  If one
restricts the analysis to six-membered rings, the most used parametrization of
puckering coordinates is probably represented by the Cremer--Pople polar coordinates
set ($Q$,$\theta$,$\phi$), that spans the configuration space using a radial
coordinate $Q$ (the total puckering amplitude), and the zenithal and azimuthal
angles $\theta$ and $\phi$, respectively. Among the other possible parametrizations,
the Cartesian one undoubtedly has some advantages\cite{Strauss70,Cremer75}.

Cremer--Pople puckering coordinates are surely the correct tool to map the conformational
space of cyclic structures, but in order to understand the properties of these
systems  one needs to know the associated free energy landscape.  Since the free
energy differences between conformers (and the barriers in between) are usually
rather large, standard computational approaches like Molecular Dynamics are ineffective. Methods exist, such as
umbrella sampling\cite{Torrie77,Bartels97} or metadynamics\cite{Laio02,Bussi06}, that allow
accelerated sampling of different conformations by adding bias forces.
In general, these accelerated sampling methods are based on the choice of a (usually
small) number of collective variables (CVs). The choice of CVs is of particular
importance for the proper reconstruction of free energy landscapes: their number
should be small in order to speed up the sampling consistently, and they should
represent every slowly evolving degree of freedom of the system. Otherwise, the estimate of
the profile could be severely biased\cite{Laio08}.  In this work we show that only
the Cremer--Pople polar coordinates are suitable
for use as CVs to perform accelerated sampling, while other parametrizations, such
as the Cartesian one, introduce strong biases in the reconstruction of the free
energy profile. As a practical example, the puckering free energy profile of the
glucuronic acid, described using a classical force-field is calculated using
two different set of CVs (polar and Cartesian). The problems arising from the use of
Cartesian coordinates are analyzed and discussed.

\section{Puckering Coordinates as Collective Variables\label{sec:CVs}}
Given the value of the atomic distances $z_j$ 
from the mean plane of the ring, the $N-3$ independent puckering coordinates are obtained as follows
(for the complete derivation see the Appendix and Ref.\cite{Cremer75}):
\begin{equation}
q_m \cos \phi_m =  \sqrt{\frac{2}{N}} \sum_{j=1}^N z_j \cos\left[\frac{2\pi}{N}m(j-1)\right] \label{eq:qcosphi}
\end{equation}
\begin{equation}
q_m \sin \phi_m = -\sqrt{\frac{2}{N}} \sum_{j=1}^N z_j \sin\left[\frac{2\pi}{N}m(j-1)\right] \label{eq:qsinphi}
\end{equation}
\begin{equation}
q_{N/2} = \sqrt{\frac{1}{N}} \sum_{j=1}^N (-1)^{j-1} z_j \mbox{ ($N$ even)},\label{eq:qeven}
\end{equation}
where the index $m$ runs from 2 to $(N-1)/2$ and to \mbox{$N/2-1$} for odd and even $N$, respectively.
The normalization factors in Eqs.(\ref{eq:qcosphi}-\ref{eq:qeven})
are such that the \emph{total puckering amplitude} $Q$ can be defined as
\begin{equation}
Q^2\equiv\sum_{j=1}^N z_j^2=\sum_m q^2_m
\label{eq:Q}
\end{equation}
for both the even and odd cases.

For a six-membered ring there is only one amplitude-phase pair $(q_2,\phi_2)$ and a single puckering coordinate $q_3$. The polar coordinate set ($Q,\theta,\phi$) and the Cartesian one ($q_x,q_y,q_z$) are related to the original ($q_2,\phi_2,q_3$) coordinate set as 
\begin{displaymath} 
\left\{ 
\begin{array}{l} 
q_2\cos\phi_2=Q\sin\theta\cos\phi = q_x\nonumber\smallskip\\
q_2\sin\phi_2=Q\sin\theta\sin\phi = q_y\nonumber\smallskip\\
q_3=Q\cos\theta = q_z.\nonumber
\end{array}
\right.
\end{displaymath}

It is worth noting, that while Cremers-Pople coordinates (or equivalent
ones) are the only rigorous coordinates to map the conformation space of puckered
rings, often a set of dihedral angles has been employed
instead(see for example \cite{Appell04}).  To our knowledge, Cremers-Pople coordinates
have been employed as CVs to sample the puckering free energy landscape only in a
recent work by Biarn\'es and coworkers\cite{Biarnes07}, where the authors employed
the Cartesian parametrization, though with an immaterial opposite sign for the
phase $\phi$.

Let us now turn to the specific problem of using Cremer--Pople coordinates as CVs
for metadynamics. Cyclic compounds are usually characterized by the presence
of a number of metastable conformations, which are generally separated by high
free energy barriers, and a proper sampling of the conformational free energy
landscape can be achieved only using some accelerated methods. A modern and efficient
class of adaptive algorithms is represented by metadynamics, in its various
flavors\cite{Laio08}.  The trait common to every metadynamics algorithm is the
presence of a history-dependent potential that drives the system out of the regions
already visited. In order for the method to be efficient, the biasing potential
has to be a function of a small number of CVs, that should also be chosen so that
the bias forces allow the system to reach every point in the CV space. This
ergodicity requirement is generally presented by saying that every slow degree of
freedom has to be represented by a CV.

In principle, for a six-membered ring structure, 3 parameters are sufficient to
map the interesting regions of the conformation space. However, to obtain a
reasonable description of a system with rigid or nearly-rigid bonds, only two
coordinates are needed.  This can be understood by noticing that the total puckering
amplitude $Q$ gives a measure of the quadratic displacement (Eq.\ref{eq:Q}) from
the middle plane.  The value of this displacement is thus bound by the finite
extension of the chemical bonds, and the density of states is therefore peaked in
a thin spherical shell of radius $Q$.  In general, only one minimum is present in
the profile of $Q$, which usually has a unimodal distribution, and is definitely
a fast degree of freedom. One can thus safely avoid using $Q$ as a CV, retaining
only the two remaining orthogonal coordinates, which, in the polar representation, are
parametrized by the angles $\theta$ and $\phi$.  The choice of the pair ($\theta$,$\phi$)
as CVs does not create any ergodicity problem, because (a) $Q$ is a fast degree
of freedom and  (b) the bias forces are always tangent to the sphere surface, thus
allowing the metadynamics to reach every point of the conformation space.

Conversely, if one wants to directly use the Cartesian coordinates 
as CVs, a number of problems arise. The bias forces
associated with $q_x$ and $q_y$ always point in a direction parallel to the
equatorial plane.  This has two simple but important consequences. 

Firstly, if one only uses two CVs, namely, $q_x$ and $q_y$, once the system
is driven to the equatorial line, the bias forces are no longer able to force
it to move along the zenithal direction. Any transition across the equatorial line
has to take place only due to real forces and thermal fluctuation.  If the free
energy landscape presents a barrier higher than the thermal energy, it will be
virtually impossible for the system to transit the equatorial line, thus rendering
the method not ergodic at the practical level.  This is why, for example, in a recent
investigation of the puckering free energy landscape of a $\beta$-glucopyranoside
performed using Car-Parrinello metadynamics\cite{Biarnes07}, Biarn\'es and coworkers
found that, during their simulation run, the system never explored the southern
hemisphere.  Moreover, the fact that the strength of the bias forces along the
$\hat{\theta}$ direction decreases as $\cos(\theta)$ means that the depth of free
energy wells and height of free energy barriers along the radial direction in the
$q_x,q_y$ plane will be systematically overestimated. The error becomes more severe
as the system approaches the equatorial line. The steep free energy barrier
at high values of $q_x^2+q_y^2$ which has been observed in \cite{Biarnes07} and
which the metadynamics was unable to surmount is precisely an artefact generated
by this biased sampling.

Secondly, since the bias forces are not only softened along the $\hat{\theta}$
direction, but are also increased along the radial one, they will start to force
the system to explore regions with values of $Q$ far from the equilibrium ones,
as soon as the system departs from the polar regions (even if the third CV, $q_z$,
is also employed). The reconstruction of the free energy profile will therefore
be unavoidably biased, by the sampling of unwanted conformations at unphysical
values of the total puckering amplitude.

We wish to stress that all these problems are not related to puckering coordinates
in general, but are due to the fact that systems of physical  interest usually
present a density of states which is concentrated in a thin spherical shell. In
order to be able to sample the entire configuration space, bias forces have to
be tangent to the sphere surface. Therefore, only linear combinations of the
unit vectors $\hat{\theta}$ and $\hat{\phi}$ are suitable to define the CV space.

\section{Simulation Methods and Results}
As a practical example, we investigated the puckering free energy profile of 
glucuronic acid, employing both ($q_x$,$q_y$) and ($\theta$,$\phi$) as CVs.  The
molecule was modeled using the classical, united-atoms, G45A4\cite{Lins05} forcefield.
The system was coupled to a thermal bath by integrating the Langevin equations of
motions with a timestep of 1 fs and a friction coefficient of 0.1 ps$^{-1}$.  We
employed the well-tempered variant of the metadynamics algorithm \cite{Barducci08},
as implemented in the {\sc grometa} simulation
package\cite{Camilloni08,Berendsen95,Lindahl01}.  The {\sc grometa} code was
modified to implement the Cartesian and polar puckering coordinates.

All simulations have been performed at a temperature of 300 K. A cut-off radius
of 1 nm has been applied for every nonbonded interaction. Gaussians were placed
in the CV space every 200 integration steps, using a starting height of 1.0 kJ/mol
and a temperature window $\Delta T=2000$K for the well-tempered algorithm.  In
addition, the width of the Gaussians has been adapted runtime by rescaling it every
1000 steps to a factor 0.2 of the root mean square distance of the associated CV,
evaluated in the last 1000-step window.  During every run, independently of which
CVs were used, the values of $Q$, $\theta$ and $\phi$ were collected, in
order to be able to investigate in parallel every possible conformational parameter.
The starting conformation was the $^4$C$_1$ chair (located approximately at
$\theta=0$) for all simulations but one, in which the ergodicity was
tested by starting the simulations from the $^1$C$_4$ inverted chair (located
at approximately $\theta=\pi$). All energies reported in the following are referred
to the global minimum.

\begin{figure}[t]
\centering
\smallskip
\includegraphics[width=.95\columnwidth]{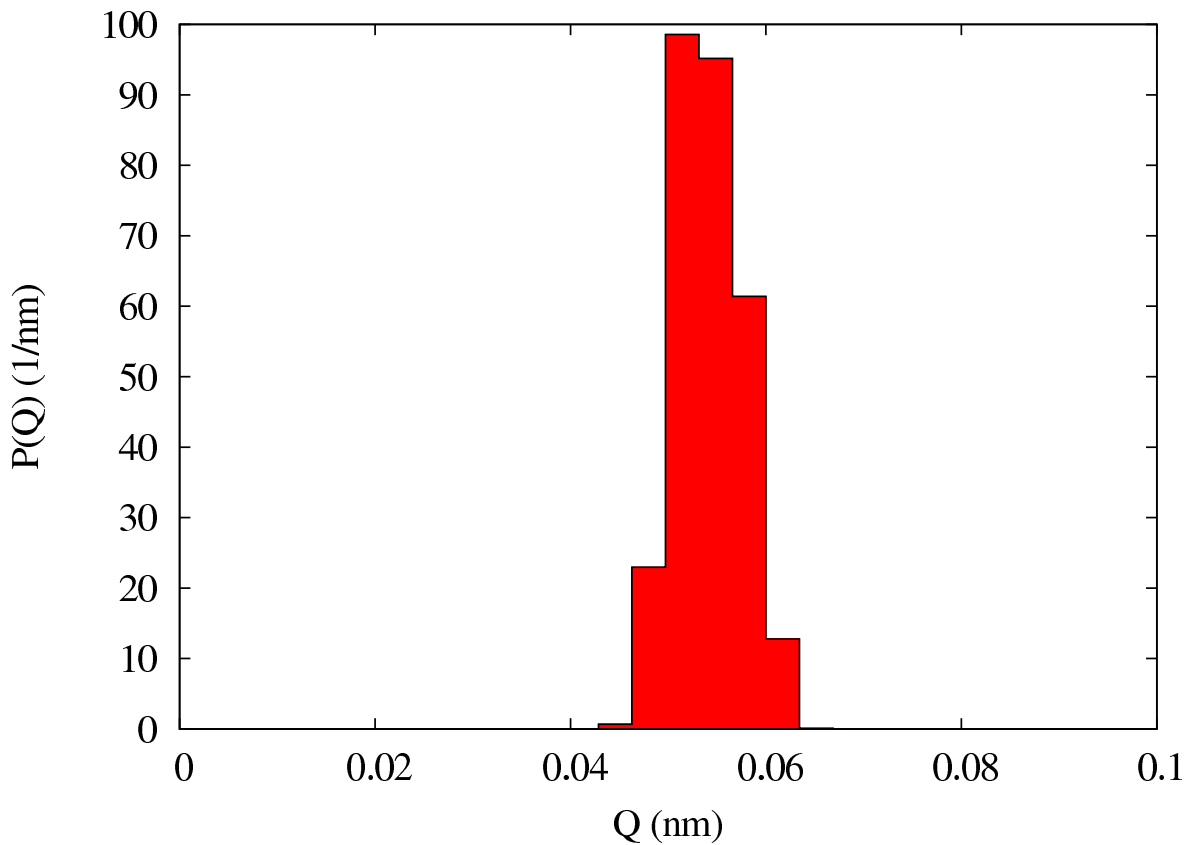}
\caption{Probability density for the radial coordinate $Q$.}
\label{fig:PQ}
\end{figure}

The first point we addressed regarded the validity of the assumption that $Q$ can
be excluded from the set of CVs needed to sample the puckered conformations free
energy landscape. In order to do this, we performed metadynamics using all three
polar CVs, and averaged out the two remaining degrees of freedom, thus obtaining
an integral profile for $Q$. The estimate for $P(Q)$, the probability distribution
of $Q$, obtained by the exponentiation of the free energy profile, is reported in
Fig.\ref{fig:PQ}.  Indeed, $P(Q)$ is characterized by a pronounced and narrow peak
located around the value of 0.05 and is unimodal, thus satisfying the requirements
needed to be considered a fast degree of freedom, and to justify the assumption
of a quasi-two-dimensional, spherical conformational space. In the following analysis,
we restricted the metadynamics to sample only the two dimensional spaces ($q_x$,$q_y$)
and ($\theta$,$\phi$).

\begin{figure}[t]
\centering
\smallskip
\includegraphics[width=1.0\columnwidth]{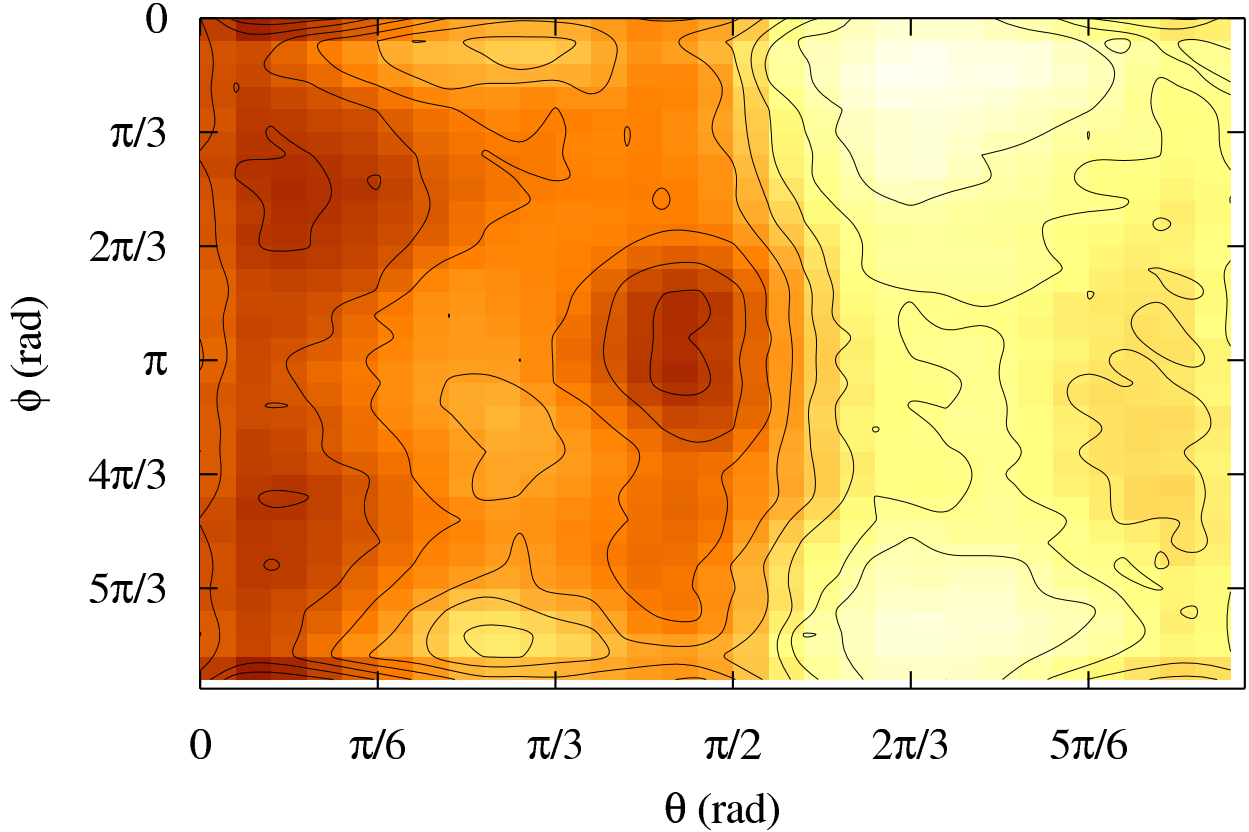}
\caption{Puckering free energy profile calculated using polar coordinates $(\theta,\phi)$. Every isoline corresponds to an increment in energy of 5 kJ/mol. (Darker colors correspond to lower energies).\label{fig:thetaphi}}
\end{figure}

The puckering free energy profile calculated using the $\theta$ and $\phi$ CVs is
presented in Fig.\ref{fig:thetaphi}. The full CV space has been spanned by the
metadynamics run, and a deep rift can be observed at $\theta\simeq0$, which contains
the global minimum corresponding to the $^4$C$_1$ conformation around $\phi=0$
and some slightly distorted chairs around $\phi=\pi/2$ and $3\pi/2$.  The next
conformer which can be observed is a $^{\mathrm{O},3}$B close to the border of the
diagram, at $\theta\simeq\pi/2$ and $\phi\simeq0$. In the middle of the diagram
($\theta\simeq\pi/2,\phi\simeq\pi$) a local minimum corresponding to the
B$_{\mathrm{O},3}$ conformer is present.  A steep barrier then has to be overome,
right after the equatorial line, to reach the $^1$C$_4$ conformation, the inverted
chair, located at around $\theta\simeq\pi$.

\begin{figure}[t]
\centering
\includegraphics[width=.51\columnwidth]{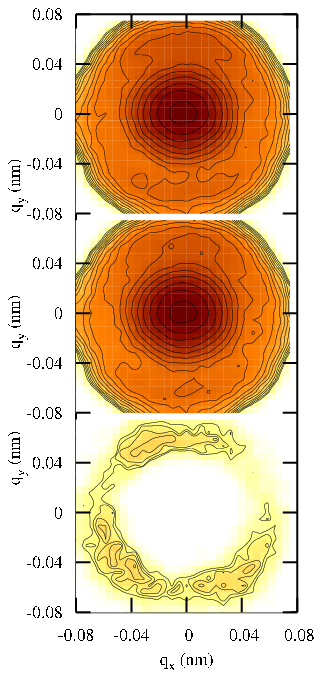}
\caption{
Top panel: free energy profile generated using the $q_x$,$q_y$ pair of CVs.
Middle and bottom panels: logarithm of the distribution of the Gaussians placed in
the northern ($G^+$) and southern ($G^-$) hemispheres, respectively, during the metadynamics run.
Every isoline corresponds to an increment in energy of 5 kJ/mol. Note that the
figures in the middle and bottom panels do not have not the meaning of a free energy;
in these cases isolines have to be understood only as a guide to the eye.
\label{fig:projection}}
\end{figure}

By employing the polar coordinates, the exploration of the southern hemisphere was
not a problem, and the barrier which has been found located close to the equatorial
line is definitely high ($\simeq 40$ kJ/mol), but easily surmountable with the
metadynamics technique.  On the contrary, the problem of ergodicity is evident if
one employs Cartesian coordinates to perform the metadynamics exploration. Since
the $q_x$ and $q_y$ coordinates alone can not distinguish between northern
and southern hemisphere, we tracked the value of $\cos(\theta)$, and were thus 
able to compute (see Fig.\ref{fig:projection}) the free energy landscape as well
as the distribution of the Gaussians placed in either of the two hemispheres. It
should be noted that the free energy $G$ is a logarithmic measure of the number
of conformations $s$ visited during the metadynamics run:  $G(s)\simeq \log\left[
\int \delta_{s,s(t')}\mathrm{d}t'\right]$. Therefore, the plots obtained by
separating out the contributions $G^\pm(s)\simeq \log\left[ \int \delta_{s,s(t')}
H(\pm\cos(\theta))\mathrm{d}t'\right]$ of the conformations in the northern and
southern hemisphere, respectively, are only indicative of the sampled regions (here
$H$ denotes the Heavside function). The two contributions $G^{\pm}$ are not to be
considered as free energies and, in particular, they do not add up to the real
free energy landscape: $G(s)\neq G^+(s)+G^-(s)$. Nevertheless, the contributions
from the two hemispheres are quite informative, and the first thing one notices
from their graphs (see Fig. \ref{fig:projection}, middle and bottom panels) is that
only the northern hemisphere has been explored completely during the metadynamics run.
This result qualitatively reproduces the behavior observed by Biarn\'es and
coworkers\cite{Biarnes07}, who noticed a sampling of the northern hemisphere only,
and attributed to the presence of an extremely high free energy barrier at
the equatorial line.

\begin{table}
\begin{tabular}{ccccc}
\hline
&$^4$C$_1$ & $^1$C$_4$ & $^{\mathrm{O},3}$B &B$_{\mathrm{O},3}$ \\
\hline
\hline
$\Delta G^\dagger$ & 0.0 & 27.7 & 9.3 & 20.0 \\
$(\theta,\phi)$ & (0.2,6.1) & (2.9,6.1)  & (1.4,0.0) & (1.5,3.4) 
\medskip\\
$\Delta G^\ddagger$ & 0.0 & n.a. & 28.7 & 33.1 \\
$(q_x,q_y)$& (0.0,0.0) & n.a. & (0.5,5.9)& (-0.5,-6.4)\\
\hline
\hline
\end{tabular}
\caption{The free energy ($\Delta G$, in kJ/mol) of different conformations,  estimated using the two different sets of Cartesian$^\dagger$ and polar$^\ddagger$ coordinates, along with the location of the conformations on the ($\theta,\phi$) plane (in rad)  and ($q_x,q_y$) plane (in $10^{-2}$nm), respectively.}
\label{table}
\end{table}

During our metadynamics run, the southern hemisphere was reached thanks to
natural fluctuations, but the sampled region is relatively small, and the metadynamics
algorithm was definitely not able to be ergodic, even though a tenfold longer simulation
time and the same Gaussian heights have been used with respect to the metadynamics
with polar coordinates.  For comparison, the estimate of the barrier height at the
equatorial line obtained using Cartesian coordinate is more than 90 kJ/mol,
compared with the value of about 40 kJ/mol estimated using polar
coordinates.  The presence of this artefact, as discussed in Section \ref{sec:CVs},
prevents the system from exploring the southern hemisphere properly and the metadynamics
from being ergodic. The magnitude of the bias which is introduced by using Cartesian
coordinates can be appreciated by looking at the differences in energy of some
selected conformations with that of the $^1$C$_4$ chair, presented in
Tab.\ref{table}.

\begin{figure}[t]
\centering
\smallskip
\includegraphics[width=1.\columnwidth]{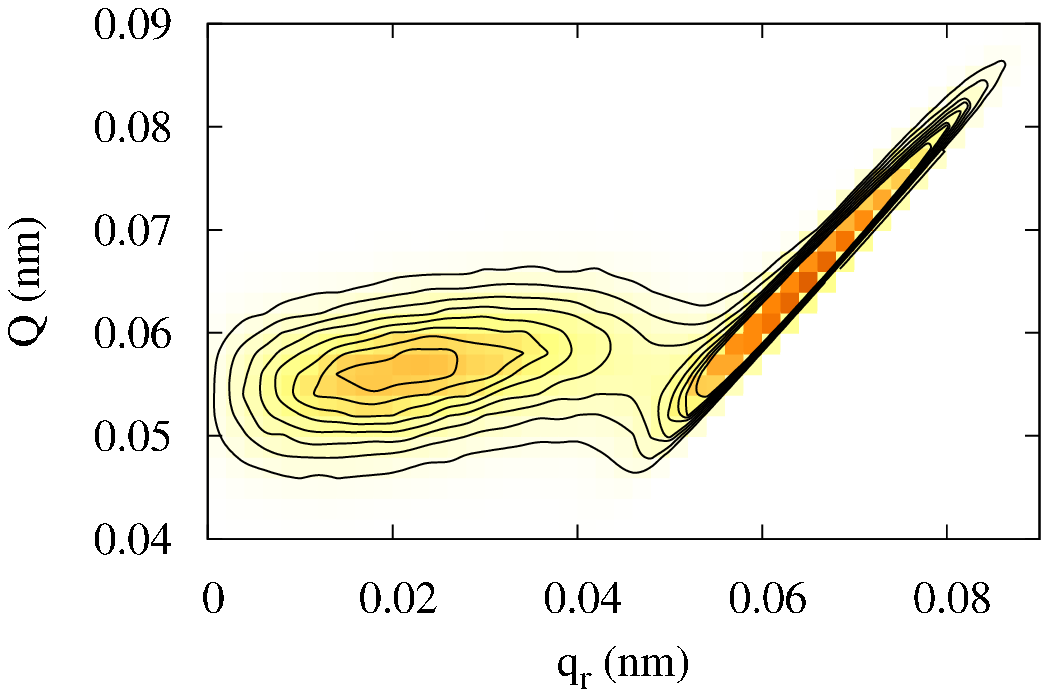}
\caption{Bivariate distribution $P(q_r,Q)$ of the projection $q_r=\sqrt{q_x^2+q_y^2}=Q\sin(\theta)$ versus the total puckering amplitude $Q$ during a metadynamics run using $q_x$ and $q_y$ as CVs. A strong correlation develops as the $q_r/Q=sin(\theta)=1$ line is approached (Lighter colors correspond to lower probabilities).\label{fig:corr}}
\end{figure}

The extent of the artefacts introduced by the use of Cartesian CVs can also be
appreciated by looking at the effect that the mixing of tangent and radial coorinates
has on the range of sampled conformations at different $Q$. In Fig.\ref{fig:corr}
the correlation between $Q$ and the magnitude of the projected puckering vector
$q_r\equiv\sqrt{q_x^2+q_y^2}$ is presented. At low values of $q_r$, that is, at
low values of the zenithal angle $\theta$, the correlation with $Q$ is minimal,
and the radial coordinate is distributed around the value of 0.055 as in the free
case. At values of $q_r$ larger than 0.05, on the other hand, a strong correlation
develops, which almost completely correlates the two variables along the
$q_r/Q=\sin(\theta)=1$ line. This means that when the system is driven close to the
equatorial line by the metadynamics, conformations that are sampled at higher
values of $\theta$ are necessarily characterized by an unphysically high total
puckering amplitude, strongly biasing the reconstructed free energy profile.

\section{Conclusions}

We have critically investigated the use of two different sets of CVs which can be
used for the study of puckered conformations of cyclic compounds.  In this
work we have shown that the Cartesian and polar sets of coordinates are
not equivalent, and that only the polar one is a proper set of CVs that allows the
ergodic and unbiased sampling of the space of different puckered conformations
for six-membered rings. We supported our conclusions by performing a metadynamics
calculation of the puckering free energy surface of a glucuronic acid ring, described
by a simple atomistic model \emph{in vacuo} using the {\sc gromos} G45A4 force-field. We showed
how the metadynamics performed using Cartesian CVs -- in contrast to the polar case
-- is non-ergodic, and how the reconstructed free energy profile suffers from strong
biases. The conformations sampled close to the equatorial line had unphysical
values of the total puckering amplitude, which has been shown to be markedly
correlated with the proximity to the equatorial line. 

In conclusion, the non-ergodicity and the biases are too high a price to be paid
for the simplifications deriving from the use of Cartesian coordinates to represent 
puckered conformations.
To the best of our knowledge, only the Cartesian variant
has been employed so far\cite{Biarnes07} as a CV set for an accelerated sampling
method, namely, metadynamics. In contrast, our analysis suggests that only
polar coordinates should be considered as a proper set of CVs when studying 
 the puckered conformations of ring structures via enhanced sampling techniques such
as metadynamics. \medskip

\section*{Acknowledgements}
The authors acknowledge the use of the Wiglaf computer cluster of the Department
of Physics of the University of Trento and thank K. Clements for critically reading
the manuscript.

\section*{Appendix: Gradients of the puckering coordinates}
For a metadynamics run it is necessary to compute gradients of the CVs,
in order to include the bias contribution to the force acting on the system.
To perform this task in this case it is necessary to recall some details in puckering coordinates construction.

The starting point are the position vectors $\mathbf{r}_j$ (here
the index $j$ runs from 1 to an arbitrary $N$) of the atoms of the ring,
with respect to an arbitrary reference frame in which we will compute derivatives.
From these vectors it is possible to calculate the geometrical center of the ring
$\mathbf{R}=\frac{1}{N}\sum_{j}\mathbf{r}_j$
and use this point as the origin of a new reference frame for the system, in which the atomic coordinate will be (summation on repeated indices is implied)
\begin{equation}
\mathbf{R}_j=\mathbf{r}_j-\mathbf{R}= \mathbf{r}_i \Delta_{ij}\;\;,\;\;\Delta_{ij}=\delta_{ij}-1/N
\label{eq:innercoord}
\end{equation}
($i=1,\ldots,N$, and $\delta_{ij}$ is the Kronecker symbol).
In order to build all puckering coordinates the projections
$z_j=\mathbf{R}_j\cdot\hat{\mathrm{n}}$  onto the 
mean plane axis $\hat{\mathrm{n}}$
are required. The mean plane axis can be univocally defined as \cite{Cremer75}
\begin{equation}
\hat{\mathrm{n}}=\frac{\mathbf{R}'\times\mathbf{R}''}{|\mathbf{R}'\times\mathbf{R}''|}
\quad,
\left\{ 
\begin{array}{l} 
\mathbf{R}'  = \sum_j \mathbf{R}_j w_{1,j}  \smallskip\\
\mathbf{R}'' = \sum_j \mathbf{R}_j v_{1,j}
\end{array}
\right.
\end{equation}
where the symbols 
\begin{equation}
w_{m,j}=\sin\frac{2\pi m(j-1)}{N} \;,\; v_{m,j}=\cos\frac{2\pi m(j-1)}{N}
\end{equation}
are defined here for further use.

The general puckering coordinate can now be rewritten as an explicit function of the atomic projections as
\begin{equation}
\left\{ 
\begin{array}{l} 
q_m    = \sqrt{2/N}\sqrt{\mathscr{A}_m^2+\mathscr{B}_m^2} \smallskip\\
\phi_m = \arctan\big[-\mathscr{A}_m/\mathscr{B}_m\big] \smallskip\\
q_{N/2}= \sqrt{1/N}\mathscr{C}\quad,
\end{array}
\right.
\end{equation}
making use of the definitions
\begin{equation}
\mathscr{A}_m=z_j w_{m,j}
\;,\;\;
\mathscr{B}_m=z_j v_{m,j}
\;,\;\;
\mathscr{C}=z_j(-1)^{j-1}
\label{eq:sumsABC}
\end{equation}
to simplify the notation.
All puckering coordinates are
eventually a function of the projections $z_j$, whose gradient is
\begin{align}
&\nabla_i z_j
=
\nabla_i \left[ \frac{\mathbf{R}_j\cdot(\mathbf{R}'\times\mathbf{R}'')}{|\mathbf{R}'\times\mathbf{R}''|} \right] \nonumber\\
&=
\frac{-z_j \nabla_i\left|\mathbf{R}'\times\mathbf{R}''\right|^2}{2|\mathbf{R}'\times\mathbf{R}''|^2}
+\frac{\nabla_i\left[\mathbf{R}_j\cdot(\mathbf{R}'\times\mathbf{R}'')\right]}{|\mathbf{R}'\times\mathbf{R}''|}.
\label{eq:gzj1}
\end{align}
Noting that $\nabla_i ( \mathbf{R}_j\cdot\hat{e}_\alpha)=\Delta_{ij}\hat{\mathrm{e}}_\alpha$ (with $\hat{e}_\alpha$ an element of the vector basis) after a somewhat long but straightforward calculation the terms of the right member in Eq.(\ref{eq:gzj1}) can be evaluated to be
\begin{displaymath}
\begin{array}{ll}
\nabla_i\left|\mathbf{R}'\times\mathbf{R}''\right|^2&
=2\Delta_{ik}w_{1,k}\big[\mathbf{R}'|\mathbf{R}''|^2-\mathbf{R}''(\mathbf{R}''\cdot\mathbf{R}') \big] + \\
&+2\Delta_{ik}v_{1,k}\big[\mathbf{R}''|\mathbf{R}'|^2-\mathbf{R}'(\mathbf{R}''\cdot\mathbf{R}') \big] 
\end{array}
\end{displaymath}
and
\begin{displaymath}
\begin{array}{ll}
\nabla_i\big[\mathbf{R}_j\cdot(\mathbf{R}'\times\mathbf{R}'')\big]
&=\Delta_{ij}\mathbf{R}'\times\mathbf{R}'' + \Delta_{ik}w_{1,k}\mathbf{R}''\times\mathbf{R}_j+\nonumber\\
&\quad +\Delta_{ik}v_{1,k}\mathbf{R}_j\times\mathbf{R}'.
\end{array}
\end{displaymath}

\bibliographystyle{apsrev}
\bibliography{biblio}

\end{document}